\documentclass[%
reprint,
superscriptaddress,
frontmatterverbose,
amsmath,amssymb,
pra,
]{revtex4-2}
 
\usepackage{amsfonts,amsmath,amssymb,amsthm}
\usepackage[normalem]{ulem}

\usepackage{graphicx}
\usepackage{dcolumn}
\usepackage{bm}

\usepackage{hyperref}

\usepackage{color}
\usepackage{amsmath}

\usepackage{lineno}
\makeatother

\newcommand{\be}{\begin{equation}}
\newcommand{\ee}{\end{equation}}

\newcommand{\bk}{{\bm{\mathrm{k}}}}
\newcommand{\br}{{\bm{\mathrm{r}}}}

\newcommand{\mR}{{\mathcal{R}}}
\newcommand{\mL}{{\mathcal{L}}}
\newcommand{\mD}{{\mathcal{D}}}
\newcommand{\mG}{{\mathcal{G}}}

\newcommand{\bx}{{\bm{\mathrm{x}}}}

\newcommand{\ii}{{\mathrm{i}}}

\usepackage{braket}
\usepackage{array}
\usepackage{booktabs}
\usepackage{multirow}

\usepackage{float}

\newif\ifdraft
\drafttrue
\draftfalse
\usepackage{titlesec}
\titleformat{\section}{\raggedright\bfseries}{\arabic{section}.}{1em}{}

\hypersetup{
colorlinks=true,
linkcolor=black
}

\begin{document}


\title{Deep learning tight-binding approach for large-scale electronic simulations \\ at finite temperatures with \textit{ab initio} accuracy}

\author{Qiangqiang Gu\normalfont\textsuperscript{$\dagger$}}

\email{guqq@aisi.ac.cn}
\affiliation{AI for Science Institute, Beijing 100080, China}
\affiliation{School of Mathematical Science, Peking University, Beijing 100871, China}

\author{Zhanghao Zhouyin}
\thanks{These authors contributed equally.}
\affiliation{AI for Science Institute, Beijing 100080, China}
\affiliation{College of Intelligence and Computing, Tianjin University, Tianjin, China, 300350}

\author{Shishir Kumar Pandey}
\affiliation{AI for Science Institute, Beijing 100080, China}
\affiliation{Birla Institute of Technology \& Science, Pilani-Dubai Campus, Dubai 345055, UAE}
\author{\\Peng Zhang}
\affiliation{College of Intelligence and Computing, Tianjin University, Tianjin, China, 300350}

\author{Linfeng Zhang}
\affiliation{AI for Science Institute, Beijing 100080, China}
\affiliation{DP Technology, Beijing 100080, China}
\author{Weinan  E}
\affiliation{AI for Science Institute, Beijing 100080, China}
\affiliation{School of Mathematical Science, Peking University, Beijing 100871, China}
\affiliation{Center for Machine Learning Research, Peking University, Beijing, 100871 China}

\date{\today}

\begin{abstract}
Simulating electronic behavior in materials and devices with realistic large system sizes remains a formidable task within the \textit{ab initio} framework due to its computational intensity. Here we show DeePTB, an efficient deep learning-based tight-binding approach with \textit{ab initio} accuracy to address this issue. By training on structural data and corresponding \textit{ab initio} eigenvalues, the DeePTB model can efficiently predict tight-binding Hamiltonians for unseen structures, enabling efficient simulations of large-size systems under external perturbations such as finite temperatures and strain. This capability is vital for semiconductor band gap engineering and materials design. When combined with molecular dynamics, DeePTB facilitates efficient and accurate finite-temperature simulations of both atomic and electronic behavior simultaneously. This is demonstrated by computing the temperature-dependent electronic properties of a gallium phosphide system with $10^6$ atoms. The availability of DeePTB bridges the gap between accuracy and scalability in electronic simulations, potentially advancing materials science and related fields by enabling large-scale electronic structure calculations.

\end{abstract}

\maketitle


Despite much progress, the simulation of materials and devices with practically relevant system sizes still remains a significant challenge within the \textit{ab initio} framework.
For instance, obtaining accurate electronic structure information with computationally expensive hybrid functionals~\cite{BeckeHybrid1993, HeydHSE2003} is crucial for band-gap engineering.
However, at the moment, performing such calculations for realistic large-scale systems is quite infeasible. 
{In addition,  atomic vibration at finite temperatures is inevitable and often influences the electronic structure.}
Thermal-driven structural changes and insulator-to-metal phase transitions, with applications in areas like thermal sensors~\cite{kimTemp2007,strelcovGas2009}, require comprehensive simulations that consider both atomic and electronic degrees of freedom.
Furthermore, the computational demands associated with accessing accurate electronic Hamiltonians for quantum transport simulations ~\cite{Brandbyge2002, Louis2003} on large systems are implausible within the current density functional theory (DFT) framework. 
Incorporating the effects of atomic vibrations in the investigation of temperature-dependent transport properties adds further complexity~\cite{liuDirect2015}.
Such simulations are beyond the capabilities of \textit{ab initio} approaches and hence often require simplified approaches that can bear the computational cost.

In this regard, the tight-binding (TB) method offers a more practical alternative for describing electronic Hamiltonians using smaller and more sparse matrices. 
Traditionally, TB Hamiltonians are constructed using empirical parameters~\cite{slaterkoster1954, Harrison1989, Tang96}, but their accuracy and transferability are often questioned. 
To address this issue, the \textit{ab initio} approach has been developed to improve the accuracy and reliability of TB models~\cite{Marzari12, AndersenNMTO2000, LuQUAMBO2004, QianQO2008}. 
It involves the projection of self-consistent \textit{ab initio} Hamiltonians onto localized bases formed by Wannier functions~\cite{Marzari12}, quasi-atomic orbital~\cite{LuQUAMBO2004, QianQO2008}, etc. 
Although one gains accuracy, the construction of the Hamiltonian remains time-consuming due to the cost associated with the \textit{ab initio} calculations and the projection step. 
Furthermore, the \textit{ab initio} TB Hamiltonian obtained this way lacks transferability to new structural configurations, limiting its applicability for electronic simulations. 
Hence, a trade-off between accuracy and efficiency is inevitable in both the traditional and {\it ab initio} TB methods. 
 
Several attempts have been made to address the dilemma of accuracy versus efficiency in modeling the electronic Hamiltonians using machine learning (ML) techniques.
Some are designed to learn the electronic Hamiltonians for molecular systems~\cite{Schuett2019,nigamEquivariant2022,fanObtaining2022}. 
For solid systems, ML approaches have been proposed to learn the Kohn-Sham (KS) Hamiltonians~\cite{HegdeBowen2017, liDeeplearning2022, zhangEquivariant2022} directly obtained from a specific DFT package based on the linear combination of atomic orbitals (LCAO) basis~\cite{Larsen2009}.
While learning the DFT Hamiltonian is straightforward, it is limited to functioning exclusively with LCAO-based DFT packages. Moreover, it is less efficient as the obtained KS Hamiltonians are way larger and denser than the TB ones with additional complexity from the overlap matrix.
Wang \textit{et. al.} \cite{wangMachine2021} designed the ML-based algorithms for generating TB matrices from electronic eigenvalues. However, no atomic structure information was considered, prohibiting its transferability to ``unseen" structures.
A subset of the authors was involved in the TBworks~\cite{guNeural2022} method, where TB Hamiltonians were constructed by learning the \textit{ab initio} eigenvalues. This approach has only been applied to one-dimensional chains.
Several other attempts are made to improve the Slater-Koster (SK) parameters from the density-functional
tight-binding framework using the ML model based on training by electronic properties such as atomic charges for molecular systems~\cite{fanObtaining2022} and by DOS for solids~\cite{sunMachine2023,mcsloyTBMaLT2023}. However, compared to the band structure, the DOS provides less insight into the electronic Hamiltonian, especially for the periodic systems. For a detailed discussion please see sections ~S12 and S13 of the Supplementary Materials (SM).
Clearly, a more general ML-based approach is warranted to efficiently generate accurate and transferable TB Hamiltonians.

In this work, we propose a deep learning-based TB method, dubbed DeePTB hereafter, to efficiently represent the electronic structure of materials with \textit{ab initio} accuracy. 
We adopt the SK framework~\cite{slaterkoster1954}, where TB Hamiltonians are constructed using gauge-invariant parameters.
DeePTB maps these parameters from symmetry-preserving local environment descriptors to obtain the TB Hamiltonian and its corresponding eigenvalues. This goes beyond the traditional two-center approximation in the empirical approaches. 
After supervised learning from training structures with \textit{ab initio} eigenvalues, DeePTB can directly predict accurate TB Hamiltonians for unseen structures during the atomic structure explorations. Such capability holds significant prominence in simulating electronic properties at finite temperatures by combining DeePTB with molecular dynamics. We further found that using eigenvalues as the training labels makes DeePTB much more flexible and independent of the choice of various bases and the form of the exchange-correlation (XC) functionals used in preparing the training labels. In addition, DeePTB can handle systems with strong spin-orbit coupling (SOC) effects. The choice of orthogonal basis and sparsity in DeePTB allows one to avoid diagonalization, enabling electronic simulations for device-size systems at finite temperatures.
All the capabilities of DeePTB extend the scope of computational science research, addressing areas that were previously difficult to explore.

\section*{Results}

We now describe the main architecture and framework of DeePTB in great detail and demonstrate its capabilities in terms of accuracy, transferability, and flexibility. We use the group-IV elemental substances and III-V group compounds as test cases. Our choice of test materials is based on the fact that they are extensively utilized in various electronic devices. 
Our method is expected to be an accurate and efficient surrogate model of DFT and applicable to a wide range of materials. 
The capability of DeePTB in dealing with realistic large-scale and long-time material simulations is demonstrated by considering a cell of one million ($10^6$) atoms of gallium phosphide (GaP) system and calculating the electronic density of states (DOS), optical conductivity, dielectric function, and refractive index at finite temperatures.

\bigskip
\noindent\textbf{Theoretical framework of DeePTB.}
The TB Hamiltonian in DeePTB takes a simplified form of the full Kohn-Sham Hamiltonian~\cite{KohnSham1965} and is based on a minimal set of localized basis functions $\left| {i, l m} \right\rangle$.
$i$ is the site index at position $\br_i$. $l$ and $m$ are angular and magnetic quantum numbers, respectively. 
The elements of TB Hamiltonian $H$ matrices can be expressed as:
\begin{equation}
H_{i,j}^{lm,l^\prime m^\prime} =  \left\langle {i,lm} \right| H \left| {j, l^\prime m^\prime} \right\rangle 
\label{eq:h_s}
\end{equation}
For $s$, $p$, and $d$ orbitals, $l$ ($l^\prime$) $=0, 1, 2$. $m$ ($m^\prime$) ranges from $-l$ to $l$. In this paper, we set the bases to be orthogonal.
The parameterization of Hamiltonian elements in Eq.~\ref{eq:h_s} takes the SK formulation~\cite{slaterkoster1954}. As for the hopping elements ($i\neq j$), it can be obtained as,
\begin{equation}
	H_{i,j}^{lm,l^\prime m^\prime} = \sum_{\zeta} \Big[ \mathcal{U}_{\zeta}(\hat{\br}_{ij}) \ h_{ll^\prime \zeta} \Big]_{mm^\prime}
	\label{eq:trans_hop}
\end{equation}
Here $\mathcal{U}_{\zeta}$ is the transformation matrix dependent solely on the direction cosines $\hat{\br}_{ij} = {\br}_{ij}/r_{ij}$ (where $r_{ij} = \lvert \br_i -\br_j \rvert$) between the two sites and  SK integrals $h_{ll^\prime{\zeta}}$ are for $\zeta$-type bond.  For instance, $p$-$p$  SK integrals include $h_{pp\sigma}$ and $h_{pp\pi}$ for $\zeta = \sigma$ and $\pi$ bonds respectively. 

As for the on-site matrix ($i=j$), a generalized strain-dependent onsite formalism by Niquet \textit{et. al.}~\cite{NiquetOnsite2009} is adopted in our work. This allows our model to simulate the strain effect, 
\begin{equation}
H_{i,i}^{lm,l^\prime m^\prime} = \epsilon_l \delta_{ll^\prime}\delta_{mm^\prime} + \sum_q \sum_{\zeta} \Big[ \mathcal{U}_{\zeta}(\hat{\br}_{iq}) \ \epsilon_{ll^\prime \zeta} \Big]_{mm^\prime}
\label{eq:trans_onsite_mat}
\end{equation}
Here, the first term takes care of only the diagonal part of onsite energies $\epsilon_l$ of different orbitals, while the second part, the strain correction term represented by  SK-like integrals $\epsilon_{ll^\prime \zeta}$, contributes when off-diagonal components of the onsite matrix are expected, for example in strained structures. $q$ runs over all the neighbors of site $i$ within a cut-off radius.
The corrections of $\varepsilon_l$ are always taken into consideration, while the second strain correction term has the flexibility to be turned off by setting $\epsilon_{ll^\prime \zeta}$ all zeros, which returns to the onsite formalism of the original SK method~\cite{slaterkoster1954}. 
To achieve better accuracy, the strain correction is included in all the reported models unless stated otherwise.

Additionally, for the systems with non-negligible SOC effect (usually the case of heavy atoms), this effect must be considered, which can be formulated as~\cite{GuSOC2023},
\begin{equation}
\hat{H}_\text{soc}=\sum_i \lambda_i \bm L_i \cdot \bm S_i 
\label{eq:H_soc}
\end{equation}
Here, $\bm L_i$ and $\bm S_i$ are the orbital and spin momentum operators with interaction strength $\lambda_i$.
The full Hamiltonian $\mathcal{H}$  with SOC effect can be constructed as $\mathcal{H}= \mathcal{I}_2 \otimes H + H_{\text{soc}}$, where $\mathcal{I}_2$ is the $2\times 2$ identity matrix and $\otimes$ is the Kronecker product. 

In short, as described in Eq.(\ref{eq:trans_hop}-\ref{eq:H_soc}), to construct the TB Hamiltonians, one needs to define the bond-wise parameters, i.e.  $h_{ll^\prime \zeta}$ and $\epsilon_{ll^\prime \zeta}$,  as well as the atomic parameters including $\epsilon_{l}$ and  $\lambda_{l}$.  
It is worth noting that in the empirical TB approach, the SK integrals are obtained based on the two-center approximation and depend only on the relative separation of the two centers, while the atomic parameters depend only on the nature of the atomic species. 
Within DeePTB, the TB parameters have not only analytical bond length dependence but also local environment-dependent corrections from neural networks. Hence, our approach is highly expressive and goes beyond the two-center approximation. The detailed NN architecture will be discussed next. 

\begin{figure}[h!]
\includegraphics[width=8.0 cm]{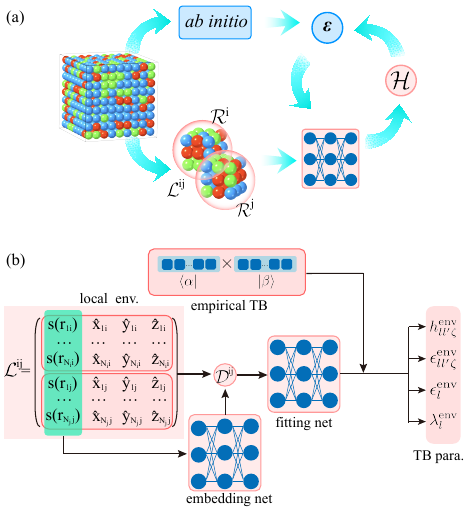}
\caption{Architecture of the DeePTB method. (a) The training workflow. $\mR^{i}$ represents the local environment of the center atom $i$, which, combined with $\mR^{j}$, forms the local environment $\mathcal{L}^{ij}$ for bond $ij$. $\mathcal{H}$ and $\bm{\varepsilon}$ denote the Hamiltonian and eigenvalues, respectively. (b) Neural network mappings from the local environment to tight-binding (TB) parameters. The scalars $s(r_{ij})$, highlighted in green, serve as inputs for the embedding network that maps $\mathcal{L}^{ij}$ to a symmetry-preserving descriptor $\mathcal{D}^{ij}$. This descriptor is then mapped to corrections to empirical TB parameters, represented by the inner products of the vectors $\langle \alpha|$ and $|\beta\rangle$, encoded by sets of neurons. The environment-dependent TB parameters, including hopping integrals $h^{\text{env}}_{ll^\prime{\zeta}}$, onsite integrals $\epsilon^{\text{env}}_{ll^\prime{\zeta}}$, onsite energies $\epsilon_l^{\text{env}}$, and spin-orbital coupling strength $\lambda_l^{\text{env}}$, are obtained.}
\label{fig:model}
\end{figure}

\bigskip
\noindent\textbf{Neural network architecture of DeePTB.}
The general architecture of DeePTB is presented in Fig.~\ref{fig:model}. The DeePTB broadly involves a three-step process. We first construct an empirical TB Hamiltonian for a system described by structure $\mR=\{\br_i\}$. Then in the second step, 
DeePTB extracts the local chemical environments, which are subsequently utilized to construct symmetry-preserving environment descriptors. These descriptors are then fed into the NN to obtain the environment-dependent TB parameters. 
In the third and final step, we train our NN model with the \textit{ab initio} electronic bands $\mathcal{E}(\mR)=\{\mathcal{E}_{n\bk}\}$ as the target. Here, $n, \bk$ are the band and lattice momentum indices respectively. Formally, these parameters can be represented as,
\begin{equation}
h^{\text{env}}_{ll^\prime{\zeta}} =  h_{ll^\prime{\zeta}}(r_{ij}) \times \left[1+\Phi_{ll^\prime\zeta}^{o_i,o_j}\left(r_{ij},\mathcal{D}^{ij}\right)\right]	
\label{eq:modelarch}
\end{equation}
For brevity, we only present the SK hopping integrals $h^{\text{env}}_{ll^\prime{\zeta}}$ as an example. Here  $r_{ij}$ is the bond length as described earlier and $\mathcal{D}^{ij}$ is the local environment descriptor of  bond-$ij$. Descriptors are invariant under the translational, permutational, and rotational symmetry operations. $\Phi_{ll^\prime\zeta}^{o_i,o_j}$ maps the $r_{ij}$ and $\mathcal{D}^{ij}$ to correct the empirical parameters $h_{ll^\prime{\zeta}}$ to provide the environment-based $h^{\text{env}}_{ll^\prime{\zeta}}$.
$o_i/o_j$ is the atomic orbital on atom $i/j$.
Other parameters can be obtained in a similar fashion, that is using environment descriptors for the correction of empirical parameters. 
We now introduce the details of each term in Eq.\eqref{eq:modelarch}.

For the empirical TB parameters terms, they are defined as analytical functions of the bond length $r_{ij}$ with several coefficients to be fitted, such as the formulas given in the literatures~\cite{slaterkoster1954, Tang96} accompanied by a cutoff function, which are available in DeePTB.
For each coefficient, we generate two vectors $\alpha, \beta\in \mathbb{R}^d$ by sets of neurons, and represent them by the inner product $\langle \alpha|\beta\rangle$. 
This way, the empirical parameters, which give a good start for the environment corrections, can be fitted efficiently with NN-based optimization when trained on the \textit{ab initio} band structures.

The local environment descriptors are constructed from the local chemical environment of each site. For site-$i$, its local chemical environment can be defined as a tensor $\mR^i\in\mathbb{R}^{N_i\times 4}$,
\begin{equation}
	(\mR^i)_q=\left(s(r_{qi}), \frac{x_{qi}}{r_{qi}}, \frac{y_{qi}}{r_{qi}}, \frac{z_{qi}}{r_{qi}}\right)
\end{equation}
where $q\in \{q|r_{qi}<r_{\mathrm{cut}}\}$ is the index of neighboring the atoms lying within a sphere of radius $r_{\mathrm{cut}}$ centered at $\br_i$.  $N_i$ is the total number of $q$ and $s(\cdot)$ is a smooth function of the scalars  $r_{qi}$.
The environment descriptor $\mD^{ij}$, inspired by DeepPot-SE model~\cite{Linfeng_NIPS_2018}, is constructed from $\mR^{i}$ and $\mR^{j}$. Here, we define bond-environment matrix as $\mathcal{L}^{ij} = (\mR^{i},\mR^{j})$ as shown in Fig.~\ref{fig:model}(b), containing the information of atomic positions. 
Similarly, we can define an embedding matrix $\mG^{ij}=(\mG^i,\mG^j)$, where each $\mG_i=\{...g_{qi},...\}$ contains the embeddings of  $s(r_{qi})$. $g_{qi}$ is mapped by a embedding neural network $G^{o_q,o_i}:\mathbb{R}\rightarrow\mathbb{R}^{M}$. Here $o_q/o_i$ denotes the chemical species of atom-$q/i$. 
Finally, the descriptor of bond-$ij$ is constructed as,
\begin{equation}
	\mD^{ij} = \frac{1}{(N_i+N_j)^2}(\mG^{ij})^T\mL^{ij}(\mL^{ij})^T\mG^{ij}_{<}
\end{equation}
where $\mG^{ij}_{<}\in \mathbb{R}^{(N_i+N_j)\times M_<}$ takes only $M_<$ columns of $\mG^{ij}$, to reduce the size of descriptors. Descriptors are proved to be invariant to the translational, permutational, and rotational symmetry operations.  The invariance under the interchange of two center atoms $ij$ to $ji$ is automatically guaranteed in $\mD^{ij}$. Therefore, only the bonds-$ij$ with $i \geq j$ need to be calculated. 

The embedding neural network $G^{o_q,o_i}$ is a function that maps the radial information of each bond to an embedding vector. The vectors are later used to construct the environmental descriptors. The fitting network $\Phi_{l l^\prime \zeta}^{o_i,o_j}$ takes environmental descriptors as input and generates TB parameters as output. $G^{o_q,o_i}$ and $\Phi_{l l^\prime \zeta}^{o_i,o_j}$ is composed of multiple layers of standard fully connected NN with optional residual connections ~\cite{he2016deep}. The fully connected NN is composed of a series of linear transformations and nonlinear activation functions. Formally, each layer is expressed as $L(\bx)=\sigma(\bm{\mathrm{W}}\bx+\bm{\mathrm{b}})$ where $\bx\in \mathbb{R}^{d_1}$ is weighted by a matrix $\bm{\mathrm{W}}\in \mathbb{R}^{d_2\times d_1}$ and biased by a vector $\bm{\mathrm{b}}\in \mathbb{R}^{d_2}$. It is then fed into the non-linear function $\sigma(\cdot)$ ($\mathrm{tanh}$ in our settings). When with residual connections, the output of layer $L$ is added with an identity mapping, giving $L(\bx)+\bx$ as the layer output. 
We employ up or downsampling techniques when the input and output dimensions mismatch.

In the last step, the obtained parameters are then transformed to the Hamiltonian matrix as defined in Eq.~\ref{eq:trans_hop}-\ref{eq:trans_onsite_mat}.
It finally leads to the environment-dependent TB Hamiltonian $H$, which is exactly diagonalized to obtain the eigenvalues. The embedding and fitting NNs are trained by minimizing the loss function defined as:
\begin{equation}
	L = \sum_{n\bk} \frac{1}{2} w_{n}
		\left[
		\lVert \mathcal{E}_{n\bk} - \hat{\mathcal{E}}_{n\bk}\rVert^2 + \lVert \Delta \mathcal{E}_{n\bk}- \Delta \hat{\mathcal{E}}_{n\bk}\rVert^2
		\right]
\end{equation}
Here, the ${\mathcal{E}}_{n\bk}$ and $\hat{\mathcal{E}}_{n\bk}$ are the eigenvalues from the predicted TB Hamiltonian and \textit{ab initio} calculations, respectively. Depending on the choice of TB basis set, the $\hat{\mathcal{E}}_{n\bk}$ is chosen from the low energy subspace of the full KS eigenvalues. The eigenvalues are sorted by their magnitude, determining the order of band-$n$, which is weighted by user-defined weights $\omega_n$.
$
\Delta {\mathcal{E}}_{n\bk} = {\mathcal{E}}_{n\bk} - {\mathcal{E}}_{n\bk^\prime}
$ 
is the difference of the eigenvalues between $\bk$ and a randomly chosen $\bk^\prime$. 
In the training process, optimization algorithms such as Adam~\cite{Adam2014} are available for the optimal parameters. The detailed training process is described in the Method section. 
Next, we proceed to showcase the performance and abilities of the DeePTB approach, considering various materials from the IV and III-V groups as examples.

\bigskip
\noindent\textbf{Validation on MD trajectories.}
\begin{figure*}[htbp!]
	\includegraphics[width=15.0 cm]{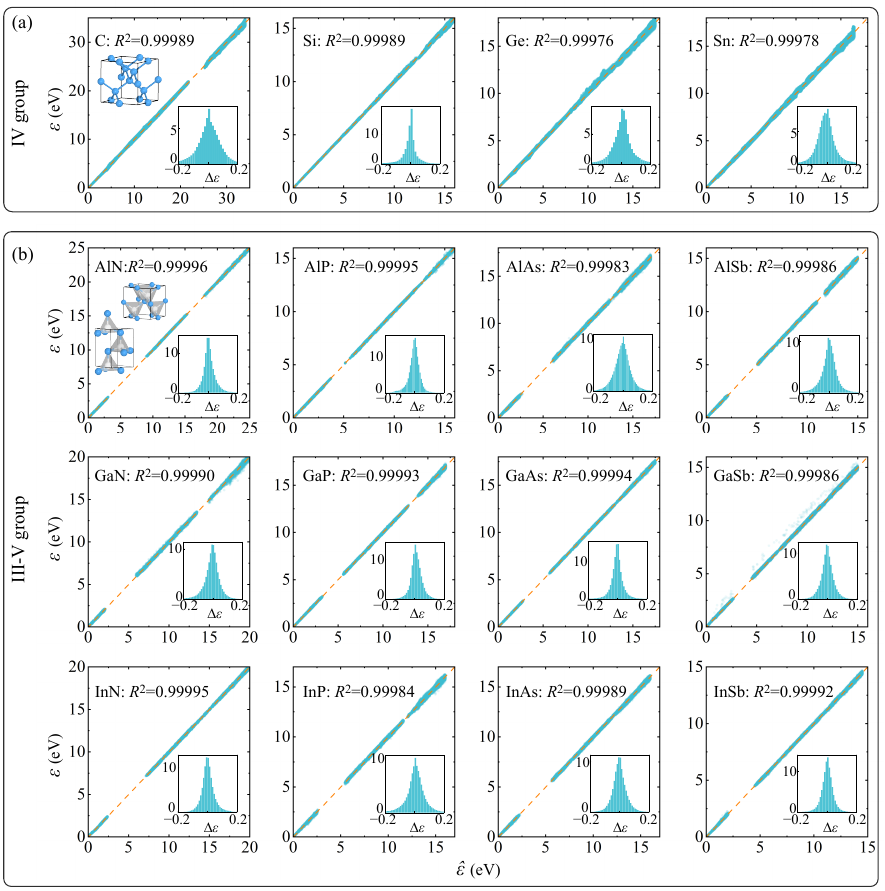}
    \caption{Validation of DeePTB predictions for group IV and III-V systems. (a) Comparison between eigenvalues from DeePTB Hamiltonians ($\varepsilon$) and those from \textit{ab initio} calculations ($\hat{\varepsilon}$) for group IV systems. (b) Comparison between eigenvalues from DeePTB Hamiltonians and those from \textit{ab initio} calculations for III-V group systems in both cubic and hexagonal phases. The minimum eigenvalue is set to zero. Insets show the distribution of eigenvalue errors ($\Delta\varepsilon = \varepsilon - \hat{\varepsilon}$) in each case, and the corresponding values of the coefficient of determination ($R^2$) are also provided. The diamond structure for group IV systems and the zincblende and wurtzite structures for III-V group systems are presented in the subplots for C and AlN systems. Source data are provided as a Source Data file.}
	\label{fig:err_hist}
\end{figure*}
We demonstrate the generalization ability of DeePTB to accurately predict TB Hamiltonians for unseen snapshots in MD trajectories. This ability is particularly valuable for electronic simulations, where the dynamic effects and interplay of ionic and electronic degrees of freedom are crucial. 
We present the test results in predicting TB Hamiltonians for unseen configurations from MD trajectories of 4 group-IV systems including diamond (C), silicon (Si), germanium (Ge), and alpha-tin ($\alpha$-Sn), as well as 12 compounds formed by group-III elements (aluminum (Al), gallium (Ga) and indium (In)) and group-V elements (nitrogen (N), phosphorus (P), arsenic (As) and antimony (Sb)).
For the group-IV systems, we focused on their cubic phase, specifically the diamond structure, as it is the most thermodynamically stable under standard conditions. Regarding the III-V group systems, we considered both the cubic phase (zincblende structure) and the hexagonal phase (wurtzite structure). This distinction is necessary because certain III-V materials are stabilized in the hexagonal phase, while others exhibit the cubic phase.

The NVT ensemble MD simulations are performed by LAMMPS package~\cite{lammps1995}  using the Tersoff potential~\cite{nakamura2000molecular, powell2007optimized} 
at a temperature $T = 300$ K with Nose-Hoover thermostat~\cite{Nose1984,Hoover1985}.
Each material was simulated within the conventional unit cell for a duration of 500 picoseconds (ps). 
Throughout the MD simulations, structure snapshots were saved every 500 femtoseconds (fs) to eliminate correlations between adjacent structures for subsequent training and testing.
The obtained structure snapshots were then used to calculate the \textit{ab initio} eigenvalues using the ABACUS package~\cite{abacus2010,abacus2016}. In these DFT calculations, we employed the double-zeta polarization (DZP) basis set and norm-conserving Vanderbilt type (ONCV)~\cite{HamannONCV2013} Perdew-Burke-Ernzerhof (PBE)~\cite{Perdew1996} functional. 
We select the first 100 snapshots of structures in each MD trajectory as the training and the last 500 snapshots as the testing data set. This choice minimizes the correlations between the training and testing data, thereby providing a robust assessment of the predictive power of the DeePTB model. More details about the MD simulations and DFT calculations can be found in the Method section.

To accurately represent the \textit{ab initio} eigenvalues from low energy subspace, DeePTB utilizes a minimal set of $spd$ orbitals as the basis and incorporates hoppings up to the 3rd nearest neighbors in the TB Hamiltonians.  We first trained the empirical TB parameters and then trained the environment-dependent corrections as described in the model architecture. The empirical TB parameters for all the testing systems are tabulated in section S15 of SM. For the local environment, the cut-off radius $r_{\mathrm{cut}}$ is also set to include 3rd nearest neighbors. For instance, in the case of Si $r_{\mathrm{cut}}=4.7$ \AA.
The DeePTB models are initially trained on the primitive unit cell to get a starting point for training on the MD-simulated configurations. 
Once training converges, the DeePTB models are employed to predict the TB Hamiltonians for unseen structures in the testing data. The accuracy of the predicted Hamiltonians is assessed by comparing their eigenvalues with their \textit{ab initio} counterparts.
The upper panel of Fig.~\ref{fig:err_hist} shows a parity plot comparing the DeePTB-predicted and \textit{ab initio} eigenvalues for structures of the group-IV materials, namely C, Si, Ge, and $\alpha$-Sn in the cubic phase. The ideal crystal structure is shown in the inset of the plot for the C system and Fig.~S1 of SM.  
The MD trajectories include distorted structures with bond length variations of approximately 10\% around the ensemble-averaged value, as depicted in Fig. S2 in the SM.
The parity plots in Fig.~\ref{fig:err_hist} for the group-IV systems exhibit an exceptional agreement, as indicated by the coefficient of determination $R^2$ of $\approx$ 0.9999. Furthermore, the mean absolute errors (MAE) for all the group-IV systems are $\sim$ 40 meV, as presented in Table~\ref{tab:MAE}. These low MAE values further reinforce the high accuracy achieved by our method.
Additionally, Fig.~S3 in the SM demonstrates that the DeePTB models successfully reproduce the primitive band structures of the group-IV systems.
All these results indicate that the DeePTB models accurately captured the underlying microscopic physics and relationships between the structural and \textit{ab initio} eigenvalues from training data.

As for III-V group systems,  the ideal zincblende and wurtzite structures are shown in the inset of the plot for AlN in Fig.~\ref{fig:err_hist} and Fig.~S1 in SM.
Despite their similar local tetrahedral structure in both phases, their band structures differ due to distinct lattice symmetries. It poses challenges for the transferability of cubic 1st nearest neighbor empirical $spds^*$ TB models to hexagonal structures, calling for separate parameterizations or more sophisticated models for different phases.
Considering this particular case, we demonstrate the capability of the DeePTB model which can handle both cubic and hexagonal phases simultaneously.
We start with training and testing our accurate model only on the cubic ($c$) phase with MAE of about 20 $\sim$ 30 meV, as shown in the $cc$ column of Table~~\ref{tab:MAE}. 
Subsequently, we utilized the cubic DeePTB models as a starting point and trained them further on mixed ($m$) data from both cubic and hexagonal ($h$) phases. 
The resulting MAE values of the obtained DeePTB models for both phases range approximately $20\sim50$ meV, as one can see in the detailed values in Table~\ref{tab:MAE} in SM.
The lower panel of Fig.~\ref{fig:err_hist} illustrates the predicted eigenvalues against their \textit{ab initio} counterparts for structures from both phases, exhibiting $R^2$ $\approx$ 0.9999. This confirms the high accuracy and reliability of the DeePTB model in capturing the underlying physics and accurately predicting eigenvalues for structures in both cubic and hexagonal phases.
Besides, Fig.~S3 and S4 in the SM demonstrate that the DeePTB model faithfully reproduces the primitive band structures of both phases, further affirming its fidelity and competence. 
Additionally, in Fig.~S8 of SM, for GaP, we show the generalization ability of DeePTB to structures where local perturbations are larger enough to cause changes in coordination numbers. We show even for the non-uniform local environment, the DeePTB model is still able to accurately predict the band structure.
To summarize this part, we demonstrated the high competence and versatility of the DeePTB approach in predicting electronic structures of different phases structures with local perturbations.

\begin{table}[!ht]
	\centering
	\tabcolsep=0.6 cm
	\caption{The mean absolute errors (MAE) in the unit of eV for IV and III-V group materials. The notations $cc/mc/mh$ in the table represent the training data types: cubic ($c$) and mixed ($m$), and the testing data types: cubic ($c$) and hexagonal ($h$)}
 \begin{tabular}{cccc}
		\hline \hline
		\specialrule{0em}{0pt}{3pt}
		\multirow{2}*{Systems} & \multicolumn{3}{c}{MAE}\\
		~ &  $cc$  &  $mc$ & $mh$ \\
		\specialrule{0em}{0pt}{3pt}
		\hline		
		\specialrule{0em}{0pt}{3pt}
		C  & 0.048 & -& - \\
		Si & 0.031 &- & - \\
		Ge & 0.044 & - & - \\
		Sn & 0.045 & - & - \\
		\specialrule{0em}{0pt}{3pt}
		\hline
		\specialrule{0em}{0pt}{3pt}
		AlN  & 0.020  & 0.021  & 0.031 \\ 
		AlP  & 0.026  & 0.028  & 0.028 \\
		AlAs & 0.033  & 0.037  & 0.054 \\
		AlSb & 0.029  & 0.036  & 0.037 \\
		GaN  & 0.022  & 0.024  & 0.048 \\
		GaP  & 0.016  & 0.019  & 0.036 \\
		GaAs & 0.029  & 0.022  & 0.030 \\
		GaSb & 0.027  & 0.031  & 0.046 \\
		InN  & 0.021  & 0.022  & 0.038 \\
		InP  & 0.027  & 0.028  & 0.053 \\
		InAs & 0.028  & 0.032  & 0.039 \\
		InSb & 0.027  & 0.024  & 0.028 \\		
		\specialrule{0em}{0pt}{3pt}
		\hline \hline
	\end{tabular}
	\label{tab:MAE}
\end{table}

\bigskip
\noindent\textbf{Validation on the larger-size and strained structures.}
Above, we explored the generalization capability of the DeePTB approach to unseen structures of the same size as the training set. 
The design of TB parameters in the DeePTB architecture, which depends on the local environment, enables its straightforward transferability to larger-size structures.
\begin{figure}[!h]
    \includegraphics[width=8 cm]{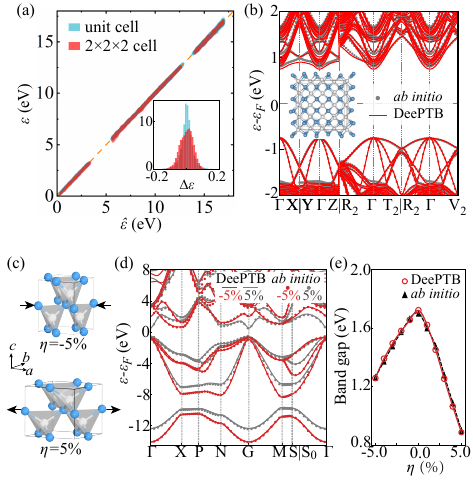}
    \caption{Generalization of DeePTB to larger-size and strained structures. (a) Comparison between eigenvalues from DeePTB ($\varepsilon$) and those from \textit{ab initio} ($\hat{\varepsilon}$) calculations of GaP for structures with $2\times2\times2$ supercell (red) and unit cell (blue). Inset: the distribution of eigenvalue errors ($\Delta\varepsilon = \varepsilon - \hat{\varepsilon}$). (b) Comparison of band structures relative to the Fermi energy ($\varepsilon_{F}$) between DeePTB (red lines) and \textit{ab initio} calculations (grey dots) in the energy window of (-2, 2) eV for the structure shown in the inset. (c) The strained structures of the GaP system with a strain of $\eta = \pm5\%$. (d) Comparison of band structures for strained GaP at $-5\%$ (red) and $5\%$ (grey) strains between DeePTB (solid lines) and \textit{ab initio} (dots) calculations. (e) Band gaps for the strained structures with $\eta$ in the range of $-5\%$ to $5\%$ from DeePTB and \textit{ab initio} calculations. Source data are provided as a Source Data file.}
\label{fig:largecell_strain}
\end{figure}
To demonstrate this generalization to larger-size structures, we consider the example of cubic GaP. We extract 500 testing structure snapshots from the MD trajectory of a $2\times2\times2$ supercell simulation box under the NVT ensemble at a temperature of $T=300$ K. This supercell dataset is then used to validate the DeePTB model, which was previously trained on the conventional smaller unit cell.
The eigenvalues obtained from DeePTB and \textit{ab initio} calculations for the $2\times2\times2$ supercell structures are plotted on top of the conventional unit cell in Fig.~\ref{fig:largecell_strain}(a). Clearly, the DeePTB model exhibits excellent agreement with \textit{ab initio} eigenvalues in both cases. In the $2\times2\times2$ supercell case, the $R^2\approx0.99996$ and testing MAE is $\approx$ 26 meV, which is only slightly larger than $19$ meV achieved in the conventional unit cell case.
The inset of Fig.~\ref{fig:largecell_strain}(b) displays a distorted structure from the testing data, where the bond length varies up to 10\% w.r.t. the undistorted ideal structure. Its complex band structure from \textit{ab initio} calculation is well reproduced by the DeePTB as shown in Fig.~\ref{fig:largecell_strain}(b) in the energy window of (-2, 2) eV. The band structure in a larger energy window of (-15, 10) eV can be found in  Fig.~S10 of SM.
These results highlight the strength of the DeePTB approach in terms of its transferability to larger-size structures. 

In many cases, particularly for tuning electronic band structures and carrier mobility~\cite{Lee_strain_bandgap, LloydGapEngineering}, one often relies on strain engineering.
However, the underlying physical changes with strain can not be straightforwardly predicted.
To account for strain effects, we incorporated an on-site strain correction term (the second term in Eq.~\eqref{eq:trans_onsite_mat}) and also introduced the local environment within the DeePTB framework. 
Here, considering the example of cubic GaP, we further demonstrate the performance of DeePTB on strained structures. 
We apply the biaxial stress perpendicular to the $z=[001]$ direction. Owing to the orthogonal lattice vectors of GaP, only non-zero strain tensor elements are $\eta_{xx}$ and $\eta_{yy}$.
Fixing $\eta_{xx}=\eta_{yy}=\eta$, we get the strained lattice structures as illustrated in Fig.~\ref{fig:largecell_strain}(c).
We set $\eta$ to vary from $-5\%$ to $5\%$, incorporating relatively large deformations that may exceed the elastic limit of materials. 
The DeePTB model was then further trained and validated on the strained configurations.
As an explicit example, Fig.~\ref{fig:largecell_strain}(d) displays the comparison between the DeePTB predicted and \textit{ab initio} band structures for $\eta=\pm 5\%$.  
One can see the excellent agreement between these two band structures. 
Additionally, Fig.~\ref{fig:largecell_strain}(e) demonstrates that the band gaps for all strained cases ($\eta \in [-5\%, 5\%]$), are accurately reproduced by the DeePTB model, with a mean absolute error (MAE) of $\sim$ 26 meV. 
These results illustrate that the DeePTB model can successfully be generalized to larger-size structures as well as accurately capture the lattice strain effects on the electronic band structure.

\bigskip
\noindent\textbf{Flexibility to different bases, XC functionals, and SOC effect.}
Previously, we employed the LCAO basis set and PBE functional for generating training eigenvalues. 
However, it is widely recognized that the accuracy and efficiency of DFT calculations depend on several factors, such as the choice of basis sets, XC functionals, and the inclusion of SOC effects. 
Unlike the approaches that directly learn DFT Kohn-Sham Hamiltonians only on the LCAO basis, DeePTB offers the advantage of being independent of such choices. To illustrate this flexibility, we consider GaAs as an example. The example of Si is shown in Fig.~S5 of SM.
\begin{figure}[h!]
	\includegraphics[width=8.0 cm]{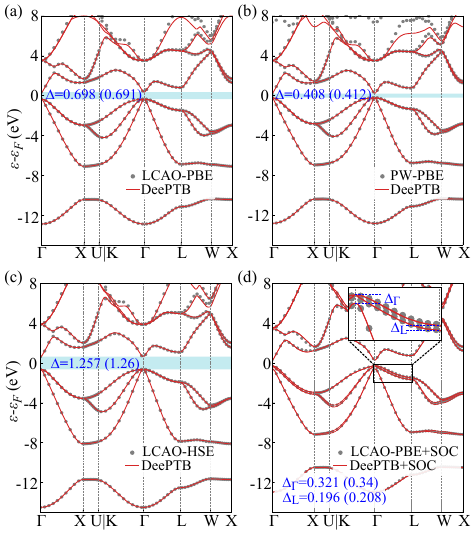}
    \caption{Comparison of band structures ($\varepsilon$) relative to the Fermi energy ($\varepsilon_{F}$) between DeePTB (red solid lines) and \textit{ab initio} (gray dots) calculations with different basis sets, exchange-correlation functionals, and inclusion of spin-orbit coupling (SOC) effects for GaAs. (a) The case for the linear combination of atomic orbitals (LCAO) basis and Perdew-Burke-Ernzerhof (PBE) functional, (b) the plane-wave (PW) basis and PBE functional, (c) LCAO basis and the Heyd-Scuseria-Ernzerhof (HSE) hybrid functional, and (d) LCAO basis with PBE functional and SOC effects. Inset: zoom-in of band structure splitting due to SOC effect at the $\Gamma$-$\mathrm{L}$ path. Shaded regions indicate band gaps ($\Delta$), whose values are displayed in the insets with units of meV. DeePTB predictions are shown alongside \textit{ab initio} calculations (values in parentheses). $\Delta_{\Gamma}$ and $\Delta_{\mathrm{L}}$ in (d) mark the energy splitting due to SOC effect at the $\Gamma$ and $\mathrm{L}$ points, respectively. Source data are provided as a Source Data file.}
\label{fig:flexibility}
\end{figure}

DeePTB models are then trained on eigenvalues of the bulk structure of GaAs from DFT calculations with different basis sets (LCAO and PW  bases) and XC functionals (PBE and Heyd-Scuseria-Ernzerhof (HSE) hybrid functional) and the exclusion/inclusion of SOC effects. 
The band structures computed from the DeePTB models are shown in Fig.~\ref{fig:flexibility}.
In the case of basis sets, we compare the LCAO and PW representations of GaAs band structures using the same PBE functional. Fig.~\ref{fig:flexibility}(a) and (b) showcase the DeePTB representation of \textit{ab initio} band structures for these different basis sets. We utilize a DZP orbital within the LCAO basis and a plane wave cutoff of 100 Ry for the PW basis. Remarkably, despite the differences in the basis sets, the DeePTB accurately reproduces the subtle variations in the band structures.
Furthermore, we investigate the impact of different XC functionals within the LCAO basis. Fig.~\ref{fig:flexibility}(a) and (c) illustrate the DeePTB band structures for GaAs calculated with the PBE and HSE functionals. The PBE functional tends to underestimate the band gap, resulting in a value of 0.691 eV for GaAs. In contrast, the HSE functional provides a more accurate band gap of 1.26 eV, much closer to the experimental value of 1.52 eV~\cite{HeydGap2005}. The DeePTB approach exhibits the ability to accurately capture the dispersion and the distinct band gaps for both XC functionals. 

As for the SOC effect, which is known to be significant for heavy atoms, we consider the cases of excluding/including SOC effects of GaAs system using the same basis and XC functionals as shown in Fig.~\ref{fig:flexibility}(a) and (d). Please refer to Fig.~S6 of SM for the cases of Sn and InSb.
Fig.~\ref{fig:flexibility}(a) and (d) demonstrate that the DeePTB band structures of GaAs agree well with the DFT-calculated ones, both with and without considering the SOC effect. The SOC effect leads to the splitting of certain energy bands, as shown in Fig.~\ref{fig:flexibility}(d). The energy splitting is most pronounced along the $\Gamma-\mathrm{L}$ path, with a splitting magnitude of $\Delta_{\Gamma}\sim0.3$ eV and $\Delta_{\mathrm{L}}\sim0.2$ eV at the $\Gamma$ and $\mathrm{L}$ points, respectively. These splittings are accurately captured by DeePTB. The inset of Fig.~\ref{fig:flexibility}(d) provides a zoomed-in view of the two split bands along the $\Gamma-\mathrm{L}$ path.

\begin{figure}[h!]
	\includegraphics[width=8.0 cm]{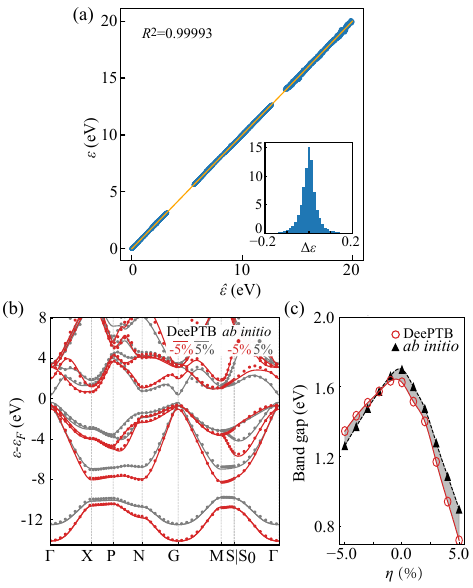}
\caption{Accuracy of DeePTB without strain correction for GaP. (a) Comparison between eigenvalues from DFT calculations ($\hat{\varepsilon}$) and DeePTB Hamiltonians ($\varepsilon$) without the strain correction term. Insets show the distribution of eigenvalue errors ($\Delta\varepsilon = \varepsilon - \hat{\varepsilon}$) and the coefficient of determination ($R^2$). (b) Comparison of band structures relative to the Fermi energy ($\varepsilon_{F}$) for strained GaP at $-5\%$ (red) and $5\%$ (grey) strains between \textit{ab initio} (dots) calculations and the DeePTB model (solid lines) without strain correction. (c) Band gaps from \textit{ab initio} and DeePTB without strain correction for the strained structures with strain $\eta$ in the range of $-5\%$ to $5\%$. The shaded area marks the difference between DeePTB predictions and \textit{ab initio} values. Source data are provided as a Source Data file.}
    \label{fig:nostrain}
\end{figure}

Till now, we have shown the application of DeePTB to elemental and binary semiconductor compounds and generalization to configurations from the MD trajectory. Now, for the sake of completeness, we demonstrate its capability on elemental copper metal (Cu), trinary semiconductor InGaAs$_2$ and carbon allotropes. As one can see in Fig.~S11(b) and (d) of SM, the metallic band structure of copper as well as the band structures of InGaAs$_2$ are well reproduced by DeePTB.  Also in Fig.~S9 of SM, one can see that a single DeePTB model can reproduce the band structures of different carbon allotropes, such as diamond and graphene.

To address the issue of fitting robustness, we perform data partitioning and random seed cross-validation. The results are provided in section Fig.~S13 of SM. In these results, random seeds as well as the cross-validation bring similar MAE with a mean value of $\sim$ 30 meV and a standard deviation of several meV in both cases. This indicates that the DeePTB model is robust to data partition and random seed selection.

\bigskip
\noindent\textbf{Performance of DeePTB with different model settings.}
Above, we have demonstrated the performance of the DeePTB model with the general settings such as the inclusion of upto 3rd nearest neighbors and the onsite strain corrections ($\epsilon_{ll^\prime\zeta}$) in the fitting. 
Below, we provide insights into the performance of DeePTB w.r.t. these settings.

First, in Fig.~\ref{fig:nostrain}(a), we show the result of DeePTB trained with excluding the strain correction term (2nd term in equation (3)), while all the other settings as well as the training and testing data are the same as that of Fig.~\ref{fig:err_hist} for GaP. The resulting MAE in this case is 29.1 meV which is 10 meV higher than the original case of 19 meV. This indicates that without the strain correction, though the model is still reasonably good, it shows a slightly higher error. 
In Fig.~\ref{fig:nostrain}(b) and (c), we calculated the band structure and band gaps for the strained structures of GaP again without strain correction, while other model settings and data are the same as that in Fig.~\ref{fig:largecell_strain}(d) and (e). 
Though the fitting of the band structures looks reasonably good in Fig.~\ref{fig:nostrain}(b), the error of the model increased. A slightly larger MAE $\sim$ 52 meV in this case compared to the original value $\sim$ 26 meV resulted in less accurate band gaps as shown in Fig.~\ref{fig:nostrain}(c). The difference in the MAE of band gaps in the two cases viz, with and without strain corrections, is $\sim$ 81 meV. This demonstrates the importance of strain corrections in dealing with systems with large lattice deformations.

\begin{figure}[h!]
	\includegraphics[width=8.0 cm]{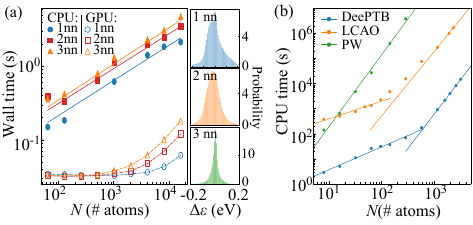}
    \caption{Computational efficiency of DeePTB. (a) Left panel: CPU and GPU wall time against the system size in log scale for DeePTB predicting Hamiltonians for Si with different nearest neighbors (nn). Right panel: the error distribution with mean absolute error values of 70, 57, and 31 meV for the DeePTB models considering up to 1st, 2nd, and 3rd nn respectively. The total number of samples for the distribution consists of 1,410,336 eigenvalues derived from 996 structures, each with 59 $\bk$-points and 24 bands. (b) Computational cost for calculating the electronic eigenvalues at the $\Gamma$ point of GaP vs. system size in \textit{ab initio} and DeePTB calculations. The dots represent values from the calculation experiments, and the lines are mathematical fits using the formula $\log(T)= \alpha \log(N) + \beta$. $T$ is CPU time and $N$ is the total number of atoms. For the plane-wave (PW) calculations, $\alpha \sim 2.9$. For the linear combination of atomic orbitals (LCAO) and DeePTB calculations, there are two stages. For both cases, $\alpha \sim 0.8$ in the first stage and $\alpha \sim 2.6$ in the second stage. Both the LCAO and PW \textit{ab initio} calculations are performed in the ABACUS package. Source data are provided as a Source Data file.}
\label{fig:efficiency}
\end{figure}

Another parameter is the choice of the number of neighbors during fitting. 
The advantages of having fewer neighbors only appear in terms of the computational cost. However, one has to pay the price for the accuracy of the DeePTB model. 
In the left panel of Fig.~\ref{fig:efficiency}(a), we plot CPU and GPU time costs for constructing the Hamiltonian (without diagonalization) with different neighbors as a function of system size $N$ (number of atoms) for Si. The CPU time is obtained on the Intel(R) Xeon(R) Platinum 8163 CPU @ 2.50GHz node and GPU time is obtained on the NVIDIA GeForce RTX 2080 Ti.
As one can see, the CPU time varies linearly on log scale with a slope of $\sim$ 0.48. 
These linear plots for different neighbors in the left panel of Fig.~\ref{fig:efficiency}(a) show an increase in wall time as the number of neighbors but the costs still stay within the same order of magnitude. 
Given the high computational power of the GPU, the wall time of the GPU does not change until a significantly large number of atoms are involved. 
For larger systems, this wall time scales approximately as $\mathcal{O}(N)$. 
However, the accuracy of the model decreases if one only considers upto 1st or 2nd neighbors. This is shown in the right panel of Fig.~\ref{fig:efficiency}(a), where the error distribution is plotted.  The MAE values of 70, 57, and 31 meV are obtained when upto 1st, 2nd, and 3rd neighbors are considered. Clearly, the prediction error increases with the decrease in the number of neighbors. 



In the case of large-size systems, one often has to choose between the computational cost and the accuracy of calculations. With DeePTB we intend to overcome this computational bottleneck. Having provided insights about the setting-dependent performance of DeePTB, we now bring the focus on the computational efficiency in the context of obtaining the eigenvalues in DeePTB and \textit{ab initio} calculations.
We compare the CPU time costs of DeePTB considering $spd$-basis and upto 3rd neighbors with those from \textit{ab initio} calculations with PBE functionals within LCAO and PW basis. We again consider DZP orbital for LCAO basis, 100 Ry for energy cutoff in PW basis and the CPU times are obtained on the Intel(R) Xeon(R) Platinum 8163 CPU @ 2.50GHz node with 64 cores. 
As shown in Fig.~\ref{fig:efficiency}(b), for smaller-size systems, both the DeePTB and LCAO \textit{ab initio} approach scale linearly with the number of atoms $N$. 
This is primarily because, for smaller-sized systems, the majority of computational costs are expended on the Hamiltonian construction procedure.
At this stage, the CPU times used by DeePTB are 2 orders of magnitude smaller than that of LCAO \textit{ab initio} calculations. 
For system size larger than $10^3$, where the diagonalization procedure 
takes the central stage in terms of computational cost, both the DeePTB and LCAO \textit{ab initio} calculations scale roughly $\sim$ $N^3$. Here,  DeePTB has the advantage of being 3 orders faster than LCAO calculations in CPU times. 
As for the PW case, it exhibits a cubic scaling behavior for all the system sizes. For the larger-size systems, DeePTB is highly efficient with $\sim$ 5 order of magnitude faster than PW calculations in terms of CPU time. 
This indicates that DeePTB provides a dramatic speedup in the simulation of electronic properties making it accessible to larger system sizes.

\begin{figure*}[ht!]
	\includegraphics[width=14.0 cm]{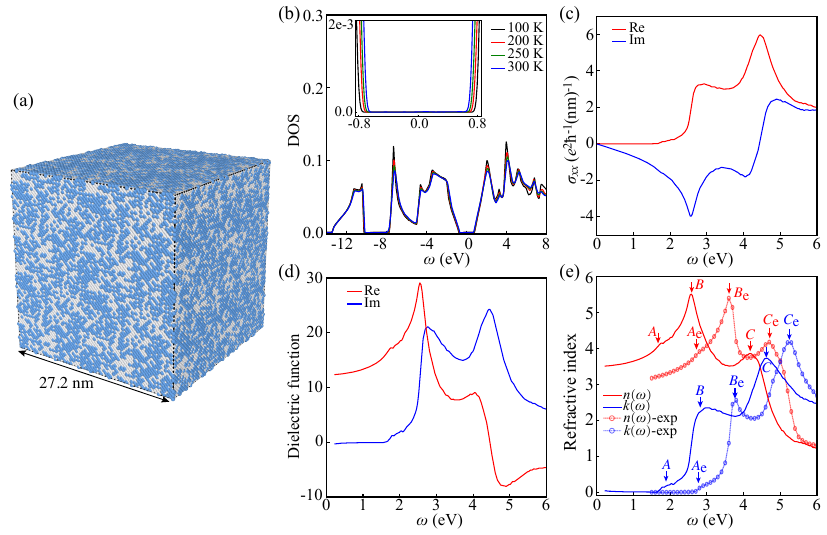}
    \caption{Simulating temperature-dependent properties for a cell of GaP with $10^6$ atoms. (a) One snapshot of the structure with $10^6$ atoms constructed from DeePMD trajectory. The blue and white balls represent Ga and P atoms, respectively. (b) Ensemble-averaged density of states (DOS) at different temperatures. Inset: zoom-in view of the DOS near the Fermi energy. (c-e) Ensemble-averaged optical properties at a temperature of 300K, including (c) optical conductivity ($\sigma_{xx}$), (d) dielectric function, and (e) complex refractive index. In (e), the experimental results from Aspnes et al.\cite{Aspnes1983} are shown for comparison. $n\left(\omega\right)$ is the real refractive index as a function of optical energy $\omega$, and $k\left(\omega\right)$ is the extinction coefficient. The peak positions marked by arrows with capital letters indicate the locations of the prominent features in the spectra. The subscript “e” denotes experimental peaks. Source data are provided as a Source Data file.}
	\label{fig:million_atoms}
\end{figure*}

\bigskip
\noindent\textbf{Application to one million atoms.}
Now that we have demonstrated the efficiency and capabilities of DeePTB, we will showcase its ability to simulate finite temperature electronic properties using an example of cubic phase GaP. To this end, we consider a large system with dimensions of $50 \times 50 \times 50$ conventional supercell containing $10^6$ atoms ($\sim 10^7$ orbitals), resulting in a simulation box with a length of approximately 27.2 nm, as illustrated in Fig.~\ref{fig:million_atoms}(a).
In such a case, the evolution and sampling of ionic configurations are performed using deep potential MD simulations powered by DeePMD-kit package~\cite{Linfeng2018, Linfeng_NIPS_2018}. 
Subsequently, DeePTB is utilized to predict the TB Hamiltonians for the structures from the obtained ionic trajectories. These Hamiltonians are then post-processed to explore electronic properties using the TB propagation method (TBPM)~\cite{Yuan2010, Yuan2011} implemented in the TBPLaS package~\cite{Yunhai2023}.

In our first demonstration, we computed the temperature-dependent electronic DOS. For a given instantaneous ionic structure $\mathcal{R}$, the corresponding electronic DOS $A_{\mathcal{R}}(\omega)$ can be calculated using  time correlation function as,
\begin{equation}
A_{\mathcal{R}}(\omega)=\frac{1}{S} \sum_{p=1}^S \frac{1}{2 \pi} \int_{-\infty}^{\infty} \mathrm{e}^{\mathrm{i} \omega t} C_{\mathcal{R}}(t) \mathrm{d}t
\end{equation}
Here, $C_{\mathcal{R}}(t)=\langle\psi_{\mathcal{R}}(0)|\psi_{\mathcal{R}}(t)\rangle$ represents the correlation function for the ionic configuration $\mathcal{R}$. $\psi(t) = \exp(-\ii H t/\hbar)\psi(0)$ denotes the time-dependent wave function, where $\hat{H}$ is the Hamiltonian operator and $\hbar$ is the reduced Planck's constant. By averaging over the MD trajectory, the temperature-dependent DOS can be obtained as $A(\omega) = \big\langle A_{\mathcal{R}}(\omega) \big\rangle_{\mathcal{R}}$, where the ionic dynamical effects on the electronic structure are intrinsically incorporated. 
Thanks to the sparsity of the DeePTB Hamiltonians, the time evolution operator can be efficiently applied to wave functions. This in turn guarantees a significant reduction in computational complexity from $\mathcal{O}(N^3)$ to nearly $\mathcal{O}(N)$ for systems with $N$-atoms. This advantage is particularly valuable when dealing with large-scale systems. 
The temperature-dependent DOS is presented in Fig.~\ref{fig:million_atoms}(b). 
Consistent with our expectation for a semiconductor, the band gap decreases as temperature increases, as illustrated in the inset of Fig.~\ref{fig:million_atoms}(b).

To substantiate our claim about the application of DeePTB, we further calculated the electronic response properties, which include optical conductivity, dielectric function, and refractive index. 
Optical conductivity, a response of the induced current density in a material to an applied optical electric field of frequency $\omega$, can be calculated using the Kubo formula~\cite{Kubo1957} as follows, 
$$
\begin{aligned}
\operatorname{Re} \sigma_{\alpha \beta}(\hbar \omega)= & \lim _{\delta \rightarrow 0^{+}} \frac{\mathrm{e}^{-\beta \hbar \omega}-1}{\hbar \omega \Omega} \int_0^{\infty} \mathrm{e}^{-\delta t} \sin (\omega t) \\
& \times 2 \operatorname{Im}\left\langle\psi\left|f(H) J_\alpha(t)[1-f(H)] J_\beta\right| \psi\right\rangle \mathrm{d} t
\end{aligned}
$$
Here, $\Omega$ represents the volume of the system, $J$ denotes the current density operator, and $f(H)$ represents the Fermi-Dirac distribution. 
Within the TBPM method, the current-current correlation enables the efficient computation of the real part of optical conductivity, denoted as $\operatorname{Re} \sigma$. 
The imaginary part, $\operatorname{Im} \sigma$, can be extracted using the Kramers-Kronig relation. 
Fig.~\ref{fig:million_atoms}(c) illustrates both the $\operatorname{Re} \sigma$ and $\operatorname{Im} \sigma$ in the energy range (0 -- 6 eV). Notably, $\operatorname{Re} \sigma$ exhibits an onset threshold corresponding to the optical band gap value at $\sim 1.5$ eV. 
After the onset threshold, $\operatorname{Re} \sigma$ increases within the energy range of 1.5 to 4.5 eV, exhibiting two peaks at 2.8 and 4.5 eV with the maximum value of $\sim 6~({e^2}/\hbar\cdot nm)$ at 4.5 eV. 
From the obtained optical conductivity, we can straightforwardly derive the dielectric function as $\sigma(\omega)=\ii\omega\epsilon_0(1-\epsilon(\omega))$, where $\epsilon_0$ represents the vacuum permittivity and $\epsilon\left(\omega\right)$ is the complex dielectric function.
The complex refractive index can be obtained from the dielectric function using
$
n^\ast\left(\omega\right)=n\left(\omega\right)+\ii k\left(\omega\right)={\epsilon\left(\omega\right)}^{1/2}
$, where $n\left(\omega\right)$ is the real refractive index  and $k\left(\omega\right)$ is extinction coefficient. 
Plots of the dielectric function and refractive index are shown in Fig.~\ref{fig:million_atoms}(d) and (e), respectively. 
In Fig.~\ref{fig:million_atoms}(e), we show a comparison of results from DeePTB and the experiments in the study of Aspnes \textit{et al}.~\cite{Aspnes1983} of $n\left(\omega\right)$ and $k\left(\omega\right)$. It can be seen that the calculated $n\left(\omega\right)$ and $k\left(\omega\right)$ exhibit identical lineshapes of the experimental counterpart, with two major peaks labeled by $B$ ($B_\text{e}$) and $C$ ($C_\text{e}$) from DeePTB (experiment).
Remarkably, even fine features like the small kink at position $A$ ($A_\text{e}$) can be well reproduced. The differences in the peak positions between the experiment and our calculations arise from the fact that GGA exchange-correlation functionals used to train the DeePTB model tend to underestimate the electronic band gap for semiconductor materials. 

\begin{figure}[ht!]
	\includegraphics[width=8.0 cm]{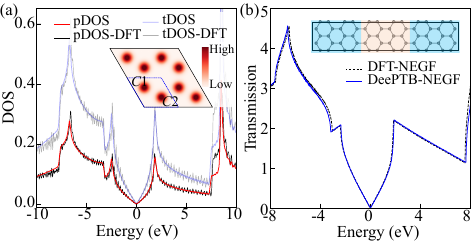}
    \caption{Density of states (DOS) and transmission coefficient calculated by DeePTB on graphene. (a) Comparison of the total DOS (tDOS) and projected DOS (pDOS) from DeePTB and density functional theory (DFT) calculations. Inset: local DOS obtained from DeePTB at energy $E_\text{0}$ = -1 eV, approximated by the Gaussian distribution centered at $C_1$ and $C_2$ sites. (b) Comparison of the transmission coefficient from DeePTB and DFT-based non-equilibrium Green's function (NEGF) calculations. Inset: the graphene device structure with left and right leads (marked in blue) and the center region (marked in yellow). Source data are provided as a Source Data file.}
	\label{fig:ldos_trans}
\end{figure}

\bigskip
\noindent\textbf{Application to the local density of states and transport properties.}
In this section, we evaluate the spatial-dependent properties to further establish our claim about the effectiveness of the DeePTB model. We performed quantum transport and local density of states (DOS) calculations on monolayer graphene, and the results are shown in Fig.~\ref{fig:ldos_trans}.
The DeePTB model using the $spd$-orbital basis and 3rd neighbors hoppings for graphene is obtained using the DFT eigenvalues as labels in the energy window of [-20, 7] eV, as shown in Fig.~S7(a) of SM. From the DeePTB model, we can get the local DOS by projecting the total DOS onto the individual atoms using the eigenvalues $\varepsilon_{n\bk}$ and eigenvectors $C_{n\bk} = (\cdots,c_{n\bk}^{v},\cdots)$, as
\begin{equation}
    A_i(E) = \sum_{v\in i} \sum_{n\bk} \delta \left(E-\varepsilon_{n\bk} \right) \left| c_{n\bk}^v\right|^2
\end{equation}
Here, $n\bk$ labels the $n$-th band at $\bk$-points in the Brillouin zone. $v$ labels the orbital basis.
$i$ is the atomic index and the total DOS is given by $A(E)=\sum_{i}{A_i(E)}$. The total DOS (tDOS) and projected DOS (pDOS) on $C$1 atom obtained from DeePTB and DFT are compared in the Fig.~\ref{fig:ldos_trans}(a). A good agreement between these two can be seen. In the inset, we show the local DOS of graphene obtained within DeePTB at $E_{\text{0}}=-1$ eV with equally distributed DOS on $C$1 and $C$2 atoms approximated by the Gaussian distribution centered at these atomic sites. We also compare the pDOS of atoms from DeePTB and DFT calculations after introducing distortion in graphene structure which breaks the symmetry of the pristine honeycomb lattice. The plot is shown in Fig.~S14 of section S14 in SM. One can see that the pDOS and tDOS from DeePTB and DFT are in excellent agreement, accurately capturing the redistribution of electronic weights among the atoms due to the introduced distortion.

In addition to the local DOS, we further perform the quantum transport simulations combining the DeePTB Hamiltonians with the non-equilibrium Green's function (NEGF) \cite{kadanoff1962} method for the graphene-based device structure shown in the inset in Fig.~\ref{fig:ldos_trans}(b). Within the NEGF formalism, the transmission coefficient is defined as, 
\begin{equation}
T(E) = \mathrm{Tr} \left[ G(E)\Gamma_L(E)G^\dagger(E)\Gamma_R(E) \right]
\end{equation}
where $G(E)$ is the retarded Green's function of the center region and $\Gamma_{L/R}$ are the broadening matrix for the left ($L$) and right ($R$) leads, respectively. With the transmission coefficient, the current-voltage (I-V) curve can be obtained as shown in Fig.~S7 of SM.
To validate our finding, we compare the transmission coefficient and I-V curve between the DeePTB and DFT-based NEGF calculations. The DFT-based NEGF calculations are performed using the TranSIESTA package~\cite{Transiesta2017}. 
A good agreement between DeePTB and DFT-based NEGF results shown in these figures demonstrates the ability of the DeePTB to provide highly reliable electronic Hamiltonians which can be further used to model quantum transport properties.

\section*{Summary and discussion}
In this work, we have introduced DeePTB, a general deep-learning-based TB approach, for predicting electronic Hamiltonians with \textit{ab initio} accuracy. 
DeePTB is designed to be independent of the choice of the basis sets and the XC functionals for generating training labels, as well as the additional capability of handling the SOC effect. 
The most compelling aspect of DeePTB is its ability to sample various electronic properties during the structural configuration simulations, and the ability to explicitly consider the effects of external perturbations like strain. We demonstrated the capabilities of DeePTB by considering the examples of III-V and IV group materials. By considering a very large GaP system of $10^6$ atoms, we substantiate our claim about the power of DeePTB by calculating temperature-dependent properties such as DOS, optical conductivity, dielectric function, and refractive index.

A few remarks about the potential of the DeePTB framework are in order. 
For different XC functionals, the dispersive features of the band structures are more or less the same. 
Therefore, one may, in principle, first train the model on computationally efficient XC functionals like LDA or GGA, and further transfer it to more costly and accurate functionals like SCAN or HSE. 
This enables the highly accurate description of experimentally observable quantities for large-scale simulations needed in cases like material simulations that are close to reality.
In order to do that, one needs efficient and accurate electronic Hamiltonians, which can be provided by DeePTB. 
The MAE obtained from DeePTB are nearly the same ($\sim$ 40 meV) as Wannier-based \textit{ab initio} TB~\cite{Marzari12}  for diamond and silicon systems (see section~S10 of SM).  Given the close values of MAE values from the two TB methods, establishes the fact the DeePTB is at the same level of accuracy as \textit{ab initio} TB methods.
Also, for large-scale samples, simulations of strain effects on electronic properties are computationally cumbersome tasks.  DeePTB can accelerate these simulations efficiently by training the model on smaller samples and transferring it to larger systems. This leads to advantages in the theoretical study of strain engineering on electronic structures. 
MD can provide the simulation of the ionic degree of freedom, that is analogous to the temperature probes of crystal structures, where ionic vibrations are a ground reality. 
DeePTB can be applied to simulate the temperature and structure-dependent electronic properties in cases where large-scale and long-time simulations are needed.
Large-scale DeePTB simulations of electronic Hamiltonian can be used to perform quantum transport simulation using techniques like NEGF. 
DeePTB makes it possible and feasible to consider other practical scenarios like defects or impurities and their influence on the electronic structure.
Another direction for DeePTB to explore is the simulation of the properties of magnetic systems. 
Given these diverse potential applications of DeePTB, we are confident that it can have far-reaching implications in the arena of electronic simulations.

\section*{Methods}
\noindent\textbf{Structure data preparation.}
The test materials contain the cubic phase group-IV systems and both cubic and hexagonal phase for III-V systems. 
We performed the MD simulations at NVT ensemble using LAMMPS~\cite{lammps1995} package based on the Tersoff force-field~\cite{nakamura2000molecular, powell2007optimized} to generate the distorted structures for training and testing. The simulation boxes are set to be the conventional unit cell, which is large enough to 3rd the nearest independent neighbors for the local environment. The lattice structures are shown in Fig. S1. 
For all the MD simulations, the temperature is set to be $300$~K using the Nose-Hoover thermostat~\cite{Nose1984, Hoover1985}.
The simulation runs for 550000 MD steps (550 ps) with a time step of 0.001 fs. The first 50 ps of simulations are thermalization. We take the snapshot configurations every 0.5 ps to get 1000 snapshots for each phase of the materials. 
We randomly select 100 configurations from the first 500 structures as the training set, while the rest 500 structures are used as the testing set to minimize the correlations between the
training and testing data.  
In order to test the generalization ability of DeePTB to a larger size, MD simulations on $2\times 2\times 2$ of conventional supercell of the GaP system are performed for further testing.  

\bigskip
\noindent\textbf{Electronic eigenvalues data preparation.}
The electronic eigenvalues data are obtained by the \textit{ab initio} calculations using the atomic-orbital based ABACUS package~\cite{abacus2010, abacus2016}. The DFT calculations are performed using LCAO bases formed by DZP orbitals with the Perdew-Burke-Ernzerhof (PBE) functional~\cite{Perdew1996} and the SG15 Optimized Norm-Conserving Vanderbilt (ONCV) pseudopotentials~\cite{HamannONCV2013}.  To demonstrate the flexibility of DeePTB with various choices of basis sets and XC functionals, we perform the DFT calculations for testing data with different basis sets (LCAO and PW bases) and XC functionals (PBE, and HSE hybrid functional).  In this case, We also consider DZP orbital within the LCAO basis, while for the PW basis, an energy cutoff of 100 Ry is used for the plane waves. In the case of SOC calculations, the fully relativistic ONCV pseudopotentials are employed.
In the self-consistent calculations, $\bk$-mesh is taken as $8\times 8 \times 8$ for the conventional unit cell and $4\times 4 \times 4$ for the  $2\times 2 \times 2$ supercell. 
The convergence threshold is set to be $10^{-7}$ for the charge density error between two sequential iterations.
The band structures are then calculated along the high-symmetry $\bk$ paths. 
After the DFT calculation, the valence bands and lower energy conduction bands are picked out, whereas the high energy empty bands and core eleven bands are ignored in the TB models. 

\bigskip
\noindent\textbf{Neural network layout and hyperparameters.}
For all models presented in our manuscript, the fitting parameters for the empirical TB parameters terms are represented by single neurons. In the embedding network, We utilize ResNet~\cite{he2016deep} to construct environment descriptors, with a neural network size of [40, 80, 160]. The fully connected network is employed for the fitting network, with a size of [200, 200, 200] for the hopping term and [100, 100, 100] for the onsite term. In cases where SOC interaction is considered, the fitting network size remains [100, 100, 100]. Tanh activation function is used across all the networks.

\bigskip
\noindent\textbf{Training process.}
The training consists of two sub-steps. First, the empirical TB parameters are initialized and trained. Here, we emphasize that the best way forward to start training is by first considering the minimal basis sets and nearest bond neighbors, which provides a rough estimation of the TB parameters. This step is physically meaningful, as it is the general scenario in solid systems where the first neighbor hoppings are the strongest. We use the minimal basis sets initially, mainly to obtain the simplest TB Hamiltonian and simultaneously to prohibit overfitting of the parameters. Then, our framework supports adding more neighbors and bases in subsequent steps and features like strain corrections and SOC effects. In the second step, we turn on the environment correction in DeePTB, which is then trained on samples from perturbed structures or MD trajectories to learn the environment dependency. The converged model can predict the TB Hamiltonians of new structures, which can be generalized to systems with different sizes and local distortions.

\bigskip
\noindent\textbf{DeePMD simulations.}
The deep potential MD simulations were performed using the LAMMPS~\cite{lammps1995} package, employing a deep learning interatomic potential (DP) model provided by DeePMD-kit~\cite{Linfeng2018, Linfeng_NIPS_2018}. The DP model was trained using the concurrent learning scheme implemented in the Deep-Potential Generator (DP-GEN). The training data set consisted of DFT calculations from the ABACUS package, employing TZDP basis sets and the PBE functional.
The DP model achieves the accuracy of a root mean square error (RMSE) value of $8.8\times 10^{-3}$ eV/atom for predicting the total energy and $2.1 \times 10^{-1}$ eV/\AA~for predicting forces.
Using the DP model, MD simulations were performed at different temperatures for 100 ps on a supercell with dimensions of $25\times25\times25$, containing $1.25\times 10^5$  atoms to generate the structure sampling for temperature-dependent electronic properties calculations.

\bigskip
\noindent\textbf{TBPM calculations.}
The TBPM~\cite{Yuan2010,Yuan2011} calculations were performed using the TBPLaS package~\cite{Yunhai2023} utilizing the DeePTB Hamiltonian to obtain electronic properties for large-size supercells of the GaP system.   The supercell structures were constructed by enlarging snapshots obtained from DeePMD simulations to a $50\times 50 \times 50$ supercell, containing a total of $10^6$ atoms.  
In TBPM, to evaluate the various time correlation functions, the random state wave functions are propagated a total of 2048 steps with a timestep of 0.105 $(\hbar/\mathrm{ eV})$.  Due to the large-size structure, only one random state sample is used to obtain the electronic properties of each structure.

\section*{Data availability}
The full data in this study are publicly available in the AI Square database (\url {https://www.aissquare.com}) under accession code DeePTBDataSet. Source data are provided with this paper.

\bigskip

\section*{Code availability}
DeePTB code\cite{dptbcode} is publicly available as an open-source package from GitHub under \url{https://github.com/deepmodeling/DeePTB}.

\section*{Competing interests}
The authors declare no competing interests.

\begin{acknowledgments}
Q.G. gratefully acknowledges fruitful discussions about the TBPLaS package with Prof. Shengjun Yuan from Wuhan University and thanks Dr. Jianchuan Liu for the deep potential model for the GaP system.  
The authors acknowledge the support of computing resources provided by the Bohrium Cloud Platform (\url{https://bohrium.dp.tech}) from DP Technology.
\end{acknowledgments}

\section*{Author contributions}

Q.G. conceived the idea and designed this work. Q.G. and Z.Z. developed the main framework of the DeePTB package. Q.G. and S.K.P. developed the SOC module in the package. Q.G., Z.Z., and S.K.P. wrote the manuscript. Q.G., Z.Z., S.K.P., P.Z., L.Z., and W.E discussed and revised the manuscript. 

%

\clearpage

\end{document}